# On calculation of quasi-two-dimensional divergence-free projections for visualization of three-dimensional incompressible flows

**Alexander Yu. Gelfgat**

School of Mechanical Engineering, Faculty of Engineering, Tel-Aviv University, Ramat Aviv, Tel-Aviv, Israel, 69978, gelfgat@tau.ac.il

**Abstract**

A visualization of three-dimensional incompressible flows by divergence-free quasi-two-dimensional projections of the velocity field on three coordinate planes was recently proposed. The projections were calculated using divergence-free Galerkin bases, which resulted in the whole procedure being complicated and CPU-time consuming. Here we propose an alternative way based on the Chorin projection combined with a SIMPLE-like iteration. The approach proposed is much easier in realization, allows for faster computations, and can be generalized for arbitrary curvilinear orthogonal coordinates. To illustrate the visualization method, examples of flow visualization in cylindrical and spherical coordinates, as well as post-processing of experimental 3D-PTV data are presented.

## 1. Introduction

This article continues our previous study on visualization of three-dimensional incompressible fluid flow [1]. There we proposed to visualize an incompressible velocity field $\boldsymbol{v} = (v_x, v_y, v_z)$ by calculating its divergence-free projections on two-dimensional coordinate planes, and computing/plotting the corresponding vector potentials which are an extension of two-dimensional stream function. For calculations of the potentials we proposed projection on the divergence free velocity bases satisfying all the boundary conditions. We have shown how these bases can be built from linear superpositions of the Chebyshev polynomials, and have illustrated the whole approach on two well-known benchmark problems: buoyancy convection in a laterally heated cube and flow in a lid-driven cubic cavity. This approach was recently implemented in [2] to describe steady and time-dependent natural convection flows.

The Chebyshev-polynomials-based approach can yield an analytic accuracy, however it is rather difficult for numerical realization and is restricted to simple domains in the Cartesian coordinates. It is also noticeably CPU-time consuming, which becomes problematic when several tens of time snapshots must be post-processed to be joined in a single animation.

In this article we propose another approach for calculation of the quasi-two-dimensional divergence free projections. This approach resembles the idea of SIMPLE iterations [5,6], where Chorin projection is applied several times within a single time step until the velocity corrections vanish. Here we reformulate this iteration procedure to make it suitable for our purposes, so that the iterations converge to one of the projections. The whole iteration procedure is relatively simple. It is argued that in the case of the standard finite volume formulation on a staggered grid, a single iteration may yield the converged result. The whole procedure is then generalized for arbitrary curvilinear orthogonal coordinates.

The proposed approach removes previous restrictions to the Cartesian coordinates and simple flow regions. We illustrate this on a problem of buoyancy convection from a hot sphere placed inside a cold cube. The visualization is carried out first in Cartesian coordinates, in which the numerical solution was computed. Then the numerical data is transformed into the spherical coordinates, where the whole visualization procedure is repeated. The latter visualization allows for a better understanding of distortion of spherical symmetry caused by the cubic walls.

For the next example we consider a classical problem of Taylor vortices appearing in the flow between rotating and stationary cylinders. We show how three-dimensional wavy vortices can be plotted in the proposed way and argue that isosurfaces of one of the potentials correspond to experimentally observed images. Finally, to show that the proposed approach can be applied also for three-dimensional experimental data, we visualize results of 3D-PTV measurements of flow in a shallow embayment.

## 2. Visualization method and numerical technique

We start with a brief description of the visualization method of [1]. A divergence-free velocity field $\boldsymbol{v} = (v_x, v_y, v_z)$ is projected on coordinate planes or surfaces in a way that (i) maintains its two-dimensional divergence equal to zero and (ii) preserves the boundary conditions. These projections are described by vector potentials that have only one non-zero component. The latter correspond to the two-dimensional stream functions in a single coordinate plane, so that its 3D isosurfaces are an extension of the two-dimensional stream function. The projected velocities are tangent to these isosurfaces.

Consider a divergence-free velocity field $\boldsymbol{v}(x, y, z)$ in an arbitrary three-dimensional flow region $V$ satisfying some boundary conditions at the region border. Considering a coordinate plane $z = const$, we define $\hat{\boldsymbol{v}}^{(z)} = (\hat{v}_x, \hat{v}_y, 0)$ to be a divergence-free projection of $\boldsymbol{v}$, if it satisfies the same boundary conditions for $v_x$ and $v_y$, and

$$\langle \boldsymbol{v} - \hat{\boldsymbol{v}}^{(z)}, \hat{\boldsymbol{v}}^{(z)} \rangle_{(x,y)} = 0, \qquad \nabla_{(x,y)} \cdot \hat{\boldsymbol{v}}^{(z)} = \frac{\partial \hat{v}_x}{\partial x} + \frac{\partial \hat{v}_y}{\partial y} = 0, \tag{1}$$

where

$$\langle \boldsymbol{v}, \boldsymbol{u} \rangle_{(x,y)} = \int_A \boldsymbol{v}(x, y, z = const) \cdot \boldsymbol{u}(x, y, z = const) dA \tag{2}$$

and $A$ is the area obtained as a cross-section of the volume $V$ by the plane $z = const$. The two-dimensional vector $\hat{\boldsymbol{v}}^{(z)}$ can be described by a stream function $\psi^{(z)}$, which, in its turn, can be interpreted as a single non-zero z-component of its vector potential $\boldsymbol{\Psi}^{(z)}$:

$$\hat{\boldsymbol{v}}^{(z)} = rot\,\boldsymbol{\Psi}^{(z)}, \quad \boldsymbol{\Psi}^{(z)} = (0, 0, \psi^{(z)}), \quad \hat{v}_x = \frac{\partial \psi^{(z)}}{\partial y}, \quad \hat{v}_y = -\frac{\partial \psi^{(z)}}{\partial x}. \tag{3}$$

Clearly, the isolines $\psi^{(z)} = const$ are streamlines of the vector $\hat{\boldsymbol{v}}^{(z)}$ in every plane $z = const$. Computing $\psi^{(z)}$ for different levels of $z$, e.g., at each grid level of a structured grid, we obtain a scalar three-dimensional function $\psi^{(z)}(x, y, z)$ that corresponds to a vector $\hat{\boldsymbol{v}}^{(z)} = (\hat{v}_x(x, y, z), \hat{v}_y(x, y, z), 0)$. The latter is called quasi-two-dimensional projection of the initial vector field $\boldsymbol{v}$. At each $z$ level, the isosurfaces of $\psi^{(z)}(x, y, z)$ are tangent to the corresponding divergent-free projection $\hat{\boldsymbol{v}}^{(z)}$. Considering similar projections in the $x = const$ and $y =$ $const$ planes, we obtain two additional scalar three-dimensional functions $\psi^{(x)}(x,y,z)$ and $\psi^{(y)}(x,y,z)$ with similar properties. Finally, three sets of isosurfaces of the three functions $\psi^{(x)}$, $\psi^{(y)}$, and $\psi^{(z)}$ complete visualization of the flow. Examples of such visualizations are given in [1].

The quasi-two-dimensional projections were calculated in [1] by projecting the initial flow field on the specially built divergence-free basis satisfying all the boundary conditions. For all the technical details the reader is referred to the above paper and references therein. This approach is straight-forward, but quite complicated and CPU-time consuming. Besides, it is limited to simple flow regions. An alternative way to calculate the quasi-two-dimensional projections is proposed below.

It was argued in [1] that application of the Helmholtz-Leray decomposition for computation of quasi-two-dimensional projections, which can be done via the Chorin projection [4], will alter boundary conditions for the velocity tangent to the boundary. However, combined with the main idea of the SIMPLE iteration [5,6], it may yield a correct and accurate result. Thus, for calculation of $\hat{\boldsymbol{v}}^{(z)}$ the following iterative procedure is proposed:

Start with $\hat{\boldsymbol{v}}^{(z)} = \boldsymbol{u}_0 = \left(v_x(x,y,z=const), v_y(x,y,z=const), 0\right)$

Repeat until $\|\varphi\| < \varepsilon$

1. Solve $\Delta_{(x,y)}\varphi = \frac{\partial^2 \varphi}{\partial x^2} + \frac{\partial^2 \varphi}{\partial y^2} = \boldsymbol{\nabla}_{(x,y)} \cdot \hat{\boldsymbol{v}}^{(z)} = \frac{\partial \hat{v}_x}{\partial x} + \frac{\partial \hat{v}_y}{\partial y}; \left[\frac{\partial \varphi}{\partial n}\right]_\Gamma = 0$.

2. Correct $\hat{\boldsymbol{v}}^{(z)}$: $\hat{\boldsymbol{v}}^{(z)} \to \hat{\boldsymbol{v}}^{(z)} - \boldsymbol{\nabla}_{(x,y)}\varphi$, i.e., $\hat{v}_x \to \hat{v}_x - \frac{\partial \varphi}{\partial x}$, $\hat{v}_y \to \hat{v}_y - \frac{\partial \varphi}{\partial y}$

3. Enforce boundary conditions for $\hat{\boldsymbol{v}}^{(z)}$ by correction of its boundary values and go to stage 1.

The above iterative procedure can be applied for any numerical discretization providing that consecutive application of approximations of the gradient and divergence operators results in the approximation of the Laplacian operator. If the iterations converge, the numerical approximation of the resulting divergence of the velocity field $\hat{\boldsymbol{v}}^{(z)}$ zeroes and the boundary conditions are satisfied. To prove that $\hat{\boldsymbol{v}}^{(z)}$ is the quasi-two-dimensional projection, it is necessary to show that the first equality in (1) holds. Indeed, this equality can be altered by stage 3 of the iterations, so that it will be satisfied only approximately. To argue that the equality holds, we observe that the correction step 2 does not alter the rotor of the field, so that if the boundary points do not pose a problem, one can expect $\nabla \times \boldsymbol{u}_0 = \nabla \times \hat{\boldsymbol{v}}^{(z)}$, which makes the difference $\left(\boldsymbol{u}_0 - \hat{\boldsymbol{v}}^{(z)}\right)$ a potential function. Then the equality $\langle \boldsymbol{v} - \hat{\boldsymbol{v}}^{(z)}, \hat{\boldsymbol{v}}^{(z)}\rangle_{(x,y)} =$ $\langle \boldsymbol{u}_0 - \widehat{\boldsymbol{v}}^{(z)}, \widehat{\boldsymbol{v}}^{(z)} \rangle_{(x,y)} = 0$ can be expected due to orthogonality of the potential and solenoidal vector fields [7].

## 3. Computation of the projection on a staggered grid

Here we assume that the three-dimensional velocity to be visualized is calculated on a staggered grid. For brevity, we define the staggered grid only in the $(x, y)$ plane and consider only one projection $\widehat{\boldsymbol{v}}^{(z)}$. Let $x_i, i = 0,1,\ldots,M$ and $y_j, j = 0,1,\ldots,N$ are arbitrarily stretched grid nodes and $x_0 < x_1 < \cdots < x_M$, $y_0 < y_1 < \cdots < y_N$. We define additionally

$$\bar{x}_{i+1/2} = \frac{1}{2}(x_i + x_{i+1}), \quad i = 0,1,\ldots,M-1; \qquad \bar{x}_{-1/2} = x_0, \ \bar{x}_{M+1/2} = x_M ; \tag{4}$$

$$\bar{y}_{j+1/2} = \frac{1}{2}(y_j + y_{j+1}), \quad j = 0,1,\ldots,N-1; \qquad \bar{y}_{-1/2} = y_0, \ \bar{y}_{N+1/2} = y_N ; \tag{5}$$

$$h^{(x)}_{i+1/2} = x_{i+1} - x_i, \quad \bar{h}^{(x)}_i = \bar{x}_{i+1/2} - \bar{x}_{i-1/2}, \tag{6}$$

$$h^{(y)}_{j+1/2} = y_{j+1} - y_j, \quad \bar{h}^{(y)}_j = \bar{y}_{j+1/2} - \bar{y}_{j-1/2} . \tag{7}$$

The scalar functions are defined in the points $(\bar{x}_{i+1/2}, \bar{y}_{j+1/2})$ and their node values are denoted by the corresponding indices, e.g., scalar potential $\varphi_{i+1/2,j+1/2}$ and $[\nabla_{(x,y)} \cdot \widehat{\boldsymbol{v}}^{(z)}]_{i+1/2,j+1/2}$. The $x$- and $y$- components of every vector are defined in the points $(x_i, \bar{y}_{j+1/2})$ and $(\bar{x}_{i+1/2}, y_j)$, respectively, and are denoted as $[\hat{v}_x]_{i,j+1/2}$ and $[\hat{v}_y]_{i+1/2,j}$. Using central finite differences we arrive to the following approximations of the gradient, divergence, and Laplacian operators ($i, j > 0$ and $i < M, j < N$)

$$\nabla_{(x,y)}\varphi = \left[\frac{\varphi_{i+1/2,j+1/2}-\varphi_{i-1/2,j+1/2}}{\bar{h}^{(x)}_i}\right]_{i,j+1/2} \boldsymbol{e}_x + \left[\frac{\varphi_{i+1/2,j+1/2}-\varphi_{i+1/2,j-1/2}}{\bar{h}^{(y)}_j}\right]_{i+1/2,j} \boldsymbol{e}_y \tag{8}$$

$$\left[\nabla_{(x,y)} \cdot \widehat{\boldsymbol{v}}^{(z)}\right]_{i+1/2,j+1/2} = \frac{[\hat{v}_x]_{i+1,j+1/2}-[\hat{v}_x]_{i,j+1/2}}{h^{(x)}_{i+1/2}} + \frac{[\hat{v}_y]_{i+1/2,j+1}-[\hat{v}_x]_{i+1/2,j}}{h^{(y)}_{j+1/2}} \tag{9}$$

$$[\Delta_{(x,y)}\varphi]_{i+1/2,j+1/2} = \frac{1}{h^{(x)}_{i+1/2}} \left\{ \frac{\varphi_{i+3/2,j+1/2} - \varphi_{i+1/2,j+1/2}}{\bar{h}^{(x)}_{i+1}} - \frac{\varphi_{i+1/2,j+1/2} - \varphi_{i-1/2,j+1/2}}{\bar{h}^{(x)}_i} \right\} +$$

$$\frac{1}{h^{(y)}_{j+1/2}} \left\{ \frac{\varphi_{i+1/2,j+3/2}-\varphi_{i+1/2,j+1/2}}{\bar{h}^{(y)}_{j+1}} - \frac{\varphi_{i+1/2,j+1/2}-\varphi_{i+1/2,j-1/2}}{\bar{h}^{(y)}_j} \right\} \tag{10}$$

Note that the above approximations are equivalent to the finite volume ones assuming linear interpolation between the grid nodes [6]. At the first step of the above iterations the problem for scalar potential $\varphi$ reduces to equality of the expressions (9) and (10) in all the inner points, and the boundary conditions are defined by zeroing one-sided finite differences at the borders

$$\varphi_{-1/2,j+1/2} = \varphi_{1/2,j+1/2}, \quad \varphi_{M+1/2,j+1/2} = \varphi_{M-1/2,j+1/2} \tag{11}$$

$$\varphi_{i+1/2,-1/2} = \varphi_{i+1/2,1/2}, \quad \varphi_{i+1/2,N+1/2} = \varphi_{i+1/2,N-1/2} \tag{12}$$

Considering the result of the correction step 3 we obtain

$$\left\{\nabla_{(x,y)} \cdot \left[\widehat{\boldsymbol{v}}^{(z)} - \nabla_{(x,y)}\varphi\right]\right\}_{i+1/2,j+1/2} = \text{Eq. (9)} - \text{Eq. (10)} = 0 \tag{13}$$

in all the inner points for $i = 1,2,\ldots,M-1$ and $j = 1,2,\ldots N-1$. If the boundary conditions are no-slip, the velocity boundary values remain unaltered. Thus, in this particular formulation the iterative process converges already after the first iteration. Clearly, it is a consequence of keeping the operator equality $div[grad\varphi] = \Delta\varphi$, or $\nabla_{(x,y)} \cdot (\nabla_{(x,y)}) = \Delta_{(x,y)}$ in the present notations, in the approximations (8)-(10). It seems that the same conclusion can be made for any other numerical method that preserves the above operator equality. However a problem may arise in the next-to-the-boundary nodes if the potential boundary values are involved in the approximation of the Laplacian operator. Here it can be easily seen that the boundary values defined by Eqs. (11) and (12) are not involved in the approximation of the gradient and Laplacian operators (8) and (10). This nice property is discussed in [6] in connection with the SIMPLE iteration technique. Here it allows us to obtain the quasi-two-dimensional projections within a single iteration. A problem of altered boundary values appears, for example, in single (not staggered) grid finite difference formulations.

Furthermore, since within the above approximation $\boldsymbol{u_0} = \widehat{\boldsymbol{v}}^{(z)} + \nabla\varphi$ after the first iteration, the first equality of (1) yields

$$\langle \boldsymbol{v} - \widehat{\boldsymbol{v}}^{(z)}, \widehat{\boldsymbol{v}}^{(z)} \rangle_{(x,y)} = \langle \boldsymbol{u_0} - \widehat{\boldsymbol{v}}^{(z)}, \widehat{\boldsymbol{v}}^{(z)} \rangle_{(x,y)} = \langle \nabla\varphi, \widehat{\boldsymbol{v}}^{(z)} \rangle_{(x,y)} = 0, \tag{14}$$

where the last equality follows from orthogonality of potential and solenoidal vector fields. To make it correct within the numerical method, the inner product has to be calculated by a quadrature formula that preserves the finite difference analog of the Gauss theorem. Therefore, within apparent numerical and accuracy restrictions, the vector $\widehat{\boldsymbol{v}}^{(z)}$ obtained via the staggered grid approximation and the single iteration is the exact quasi-two-dimensional projection of the initial velocity field.

Since visualization does not require a very good accuracy, it is possible to interpolate any experimental or numerical result on a staggered grid, and then calculate the three quasi-two-dimensional projections via a robust and straight-forward numerical procedure.

## 4. Generalization for arbitrary curvilinear orthogonal coordinates

Consider the divergence-free flow $\mathbf{v} = (v_1, v_2, v_3)$ defined in arbitrary curvilinear orthogonal coordinates $(x_1, x_2, x_3)$. The Lamé coefficients $h_i$ and the vector differential operators are defined as

$$h_i = \left(\frac{\partial x}{\partial x_i}\right)^2 + \left(\frac{\partial y}{\partial x_i}\right)^2 + \left(\frac{\partial z}{\partial x_i}\right)^2, \quad i = 1,2,3 \tag{15}$$

$$grad(\varphi) = \sum_{i=1}^{3} \frac{1}{h_i} \frac{\partial \varphi}{\partial x_i} \mathbf{e}_i \tag{16}$$

$$div(\mathbf{v}) = \frac{1}{h_1 h_2 h_3} \left[\frac{\partial}{\partial x_1}(h_2 h_3 v_1) + \frac{\partial}{\partial x_2}(h_1 h_3 v_2) + \frac{\partial}{\partial x_3}(h_1 h_2 v_3)\right] \tag{17}$$

$$rot(\mathbf{v}) = \frac{1}{h_2 h_3}\left[\frac{\partial(h_3 v_3)}{\partial x_2} - \frac{\partial(h_2 v_2)}{\partial x_3}\right]\mathbf{e}_1 + \frac{1}{h_1 h_3}\left[\frac{\partial(h_1 v_1)}{\partial x_3} - \frac{\partial(h_3 v_3)}{\partial x_1}\right]\mathbf{e}_2 + \frac{1}{h_1 h_2}\left[\frac{\partial(h_2 v_2)}{\partial x_1} - \frac{\partial(h_1 v_1)}{\partial x_2}\right]\mathbf{e}_3 \tag{18}$$

$$\Delta\varphi = div \cdot grad(\varphi) = \frac{1}{h_1 h_2 h_3}\left[\frac{\partial}{\partial x_1}\left(\frac{h_2 h_3}{h_1}\frac{\partial \varphi}{\partial x_1}\right) + \frac{\partial}{\partial x_2}\left(\frac{h_1 h_3}{h_2}\frac{\partial \varphi}{\partial x_2}\right) + \frac{\partial}{\partial x_3}\left(\frac{h_1 h_2}{h_3}\frac{\partial \varphi}{\partial x_3}\right)\right] \tag{19}$$

The divergence-free projection $\hat{\mathbf{v}}^{(3)} = (\hat{v}_1, \hat{v}_2, 0)$ of the flow $\mathbf{v}$ on a coordinate surface $(x_1, x_2, x_3 = const)$ is defined similarly to (1), (2):

$$\langle \mathbf{v} - \hat{\mathbf{v}}^{(3)}, \hat{\mathbf{v}}^{(3)} \rangle_{(x_1, x_2)} = 0, \quad \nabla_{(x_1, x_2)} \cdot \hat{\mathbf{v}}^{(z)} = \frac{1}{h_1 h_2}\left[\frac{\partial}{\partial x_1}(h_2 h_3 \hat{v}_1) + \frac{\partial}{\partial x_2}(h_1 h_3 \hat{v}_2)\right] = 0, \tag{20}$$

where

$$\langle \mathbf{v}, \mathbf{u} \rangle_{(x_1, x_2)} = \int_A \mathbf{v}(x_1, x_2, x_3 = const) \cdot \mathbf{u}(x_1, x_2, x_3 = const) dA \tag{21}$$

and $A$ is a surface obtained as a cross-section of the flow region $V$ by the coordinate surface $x_3 = const$. The two-dimensional vector $\hat{\mathbf{v}}^{(3)}$ can be described by a stream function $\psi^{(3)}$ defined as

$$\hat{v}_1 = \frac{1}{h_2 h_3}\frac{\partial(h_3 \psi^{(3)})}{\partial x_2}, \quad \hat{v}_2 = -\frac{1}{h_1 h_3}\frac{\partial(h_3 \psi^{(3)})}{\partial x_1}. \tag{22}$$

The stream function $\psi^{(3)}$ is a non-zero component of the vector potential $\mathbf{\Psi}^{(3)} = (0, 0, \psi^{(3)})$, so that $\hat{\mathbf{v}}^{(3)} = rot(\mathbf{\Psi}^{(3)})$. The two other vector potentials, $\mathbf{\Psi}^{(1)}$ and $\mathbf{\Psi}^{(2)}$ are defined in similar ways. The proposed iterative procedure is generalized as follows

Start with $\hat{\boldsymbol{v}}^{(3)} = \boldsymbol{u}_0 = (\hat{v}_1(x_1, x_2, x_3 = const), \hat{v}_2(x_1, x_2, x_3 = const), 0)$

Repeat until $\|\varphi\| < \varepsilon$

1. Solve

$$\Delta_{(x_1,x_2)}\varphi = \frac{1}{h_1 h_2}\left[\frac{\partial}{\partial x_1}\left(\frac{h_2 h_3}{h_1}\frac{\partial \varphi}{\partial x_1}\right) + \frac{\partial}{\partial x_2}\left(\frac{h_1 h_3}{h_2}\frac{\partial \varphi}{\partial x_2}\right)\right] = \boldsymbol{\nabla}_{(x_1,x_2)} \cdot \hat{\boldsymbol{v}}^{(z)} =$$
$$= \frac{1}{h_1 h_2}\left[\frac{\partial(h_2 h_3 \hat{v}_1)}{\partial x_1} + \frac{\partial(h_1 h_3 \hat{v}_2)}{\partial x_2}\right]; \qquad \left[\frac{\partial \varphi}{\partial n}\right]_\Gamma = 0.$$

2. Correct $\hat{\boldsymbol{v}}^{(z)}$: $\hat{\boldsymbol{v}}^{(z)} \to \hat{\boldsymbol{v}}^{(z)} - \boldsymbol{\nabla}_{(x_1,x_2)}\varphi$, i.e., $\hat{v}_{x_1} \to \hat{v}_{x_1} - \frac{1}{h_1}\frac{\partial \varphi}{\partial x_1}$, $\hat{v}_{x_2} \to \hat{v}_{x_2} - \frac{1}{h_2}\frac{\partial \varphi}{\partial x_2}$

3. Enforce boundary conditions for $\hat{\boldsymbol{v}}^{(z)}$ by correction of its boundary values and go to stage 1.

## 5. Examples of flows visualization

At the first stage of verification of the proposed technique of calculation of the quasi-two-dimensional velocity projections we repeated visualizations of [1] for convection and lid driven flows in a cubic box. The results were identical to those of [1]. We confirmed also that all the projections are calculated correctly in a single iteration since all the visualized results were obtained on staggered grids.

Below we show two examples of flows either computed and visualized in curvilinear orthogonal coordinates, or computed in Cartesian coordinates, but visualized in curvilinear ones after the corresponding coordinates and velocities transformations. Since the following examples have a mainly illustrative purpose, we do not focus on details of numerical simulations. At the same time, for possible comparison purposes, we report minimal, maximal and plotted level values of functions shown in the graphs. The third example illustrates visualization of 3D experimental data.

### 5.1. *Air convection from a hot ball inside a cold square box*

Consider air convection from a hot isothermal sphere placed in the center of a cubic box with cold walls. The dimensionless radius of the sphere is 0.2, and the dimensionless box length is 1. Formulation of the problem and dimensionless parameters are given in [8,9]. The numerical solutions of [9] are used for the following visualization. The computations were performed on $100^3$ and $200^3$ finite volume grids. The finite volume method was combined with the immersed boundary technique to describe the spherical boundary. Since the flows were computed on a Cartesian grid, their visualization in Cartesian coordinates is carried out first and is shown in Fig. 1 for two values of the Rayleigh number, $10^4$ and $10^6$. For computations of the quasi-two-dimensional projections, we apply the proposed iterative procedure so that no-slip boundary conditions on the sphere are enforced at the stage 3. Note that after each iteration, the staggered finite volume grid yields correct boundary conditions at the box boundaries, however, no-slip conditions at the spherical boundary are satisfied only at the end of the iteration process. The iterations were considered converged when the absolute value of the potential $\varphi$ was lesser than $10^{-8}$ pointwise. After the iterations converged, we examined the first equation of (1) using the finite volume based quadrature formula, e.g.,

$$max|\langle \boldsymbol{v} - \hat{\boldsymbol{v}}^{(z)}, \hat{\boldsymbol{v}}^{(z)} \rangle_{(x,y)}| \approx \max_k \left| \sum_{i=0}^{M} \sum_{j=0}^{N} \left\{ [(v_x - \hat{v}_x)\hat{v}_x]_{i,j+1/2,k+1/2} \bar{h}_i^{(x)} h_{j+1/2}^{(y)} + \right.\right.$$
$$\left.\left. + [(v_y - \hat{v}_y)\hat{v}_y]_{i,j+1/2,k+1/2} \bar{h}_j^{(y)} h_{i+1/2}^{(x)} \right\} \right| \quad (15)$$

The calculations showed that we needed about 120 and 80 iterations to converge in the cases of $100^3$ and $200^3$ grids, respectively. The lesser number of iterations needed for the finer grid can be explained by a shorter distance between the sphere surface and grid points, so that enforcement of the no-slip condition on the sphere lesser altered the divergence free field obtained at the second iterations stage. After iterations converged, absolute values of the integrals (15) were below $10^{-15}$, which confirms the correctness of computed projections within the numerical model.

Visualizations presented in Fig. 1 are carried out for two values of the Rayleigh number, $10^4$ and $10^6$. The potential isosurfaces are shown in color, and quasi-two-dimensional projection vectors by black arrows. It is clearly seen that these vectors are tangent to the potential isosurfaces, as expected. Since the flow is symmetric with respect to 90° rotation around the vertical axis, the potentials $\Psi^{(x)}$ and $\Psi^{(y)}$ transform one into another by this rotation. The potential $\Psi^{(z)}$ shows the fluid motion in the (*x,y*) planes that results into 8 circulating structures and represent the "horizontal" part of the convective ascending/descending flow, which is caused by viscous friction at the spherical and cubic boundaries.

To examine how the spherical symmetry is distorted by the cubic walls, we transform the result into spherical coordinates $(R, \theta, \phi)$ with the pole placed in the cube center. Then we follow the procedure of Section 4 to compute three vector potentials $\Psi^{(\phi)}$, $\Psi^{(R)}$, and $\Psi^{(\theta)}$ shown in Fig. 2. Now the no-slip conditions on the sphere are satisfied at each iteration, while the boundary conditions on the square walls must be enforced at each iteration. Note that if the whole problem was spherically symmetric (e.g., convection between two concentric spheres) one would expect to obtain a $\phi$-independent flow with $v_\phi = 0$. Then the potential $\Psi^{(\phi)}$ would also depend only on $R$ and $\theta$, and its isolines in cross-sections $\phi = const$ would coincide with the flow streamlines. When the spherical symmetry is distorted, the result becomes also $\phi$-dependent. This distortion is clearly represented by the isosurfaces of $\Psi^{(\phi)}$ (Fig. 2) that lose spherical symmetry, but remain symmetric with respect to the 90º rotation around the vertical axis, as is prescribed by the problem symmetry. Two other potentials $\Psi^{(R)}$, and $\Psi^{(\theta)}$ also preserve this symmetry (Fig. 2). They represent projections of the flow field onto spheres $R = const$ and onto cones $\theta = const$, which is illustrated additionally in Fig. 3.

Projections of the velocity vectors on the planes $\phi = 0$ and $\pi$ are plotted in Fig. 3a, where two non-zero components of the velocity projection are shown by arrows. It is clearly seen that the projected velocity is tangent to the plotted isosurface of the potential $\Psi^{(\phi)}$. This projection corresponds to the flow ascending near the hot ball and descending near the cold walls. In spite of all the results obtained in the Cartesian coordinates, one frame in Fig. 3a appears to be more informative than the two frames corresponding to the potentials $\Psi^{(x)}$ and $\Psi^{(y)}$ in Fig. 1. Comparing two frames in Fig. 3a we observe also that at $Ra = 10^4$ distortion of the spherical symmetry is similar above and below the sphere. At a noticeably larger Rayleigh number, $Ra = 10^6$, the shape of the isosurface below and above the sphere is different. Another interesting observation is the following. At $Ra = 10^4$, the wider part of the isosurface is located in front of the cube corner, while its narrower part is placed in front of the sidewall. It becomes vice versa at $Ra = 10^6$, where we find the narrower part is in front of the corner, while the wider part is in front of the sidewall.

Figures 3b and 3c provide an additional insight on how the spherical symmetry is destroyed. Absence of the spherical symmetry at the outer boundaries leads to the appearance of a circulating motion, as it was already observed in the horizontal planes in Fig. 1. This circulating motion is decomposed into one located on the spheres $R = const$ (Fig. 3b) and the other one located on the cones $\theta = const$. Again we can point to interesting changes in the potential and velocity projections patterns that take place with the increasing Rayleigh number. Thus, at a relatively low Rayleigh number, $Ra = 10^4$, the isosurfaces of $\Psi^{(R)}$ lay on the coordinate surface $R = const$ (Fig. 3b). At larger $Ra = 10^6$, the lower parts of the $\Psi^{(R)}$ isosurfaces are shifted "inside" the coordinate surface, on which we observe two recirculations instead of one. Change of the $\Psi^{(\theta)}$ pattern is even more drastic (Fig. 3c). At $Ra = 10^4$ we observe positive (red) and negative (blue) valued isosurfaces located one above another. The corresponding recirculating motion direction is opposite on positive and negative isosurfaces. At $Ra = 10^6$ the patterns of positive and negative isosurfaces change qualitatively, so that at a fixed polar angle $\phi$, we observe either positive or negative values of the potential $\Psi^{(\theta)}$. Thus we notice that at $\phi = const$ cross-sections of the right frame of Fig. 3c, the recirculating motion has the same direction along all values of $R$ and $\theta$.

*5.2. Axisymmetric and wavy Taylor vortices*

Consider the classic Taylor-Couette problem between a rotating inner and stationary outer cylinder [10,11]. The cylinders are assumed to be coaxial and infinite in the axial direction. The problem is considered in the cylindrical coordinates $(r, \theta, z)$. The cylinder radii are denoted as $R_{in}$ and $R_{out}$, the gap is $d = R_{out} - R_{in}$, and the radii ratio is $\eta = R_{in}/R_{out}$. The angular velocity of the inner cylinder is $\Omega$, and kinematic viscosity of the fluid is $\nu$. Following [11], the problem is rendered dimensionless by choosing $d$, $d^2/\nu$ and $\Omega d$ as scales of length, time and velocity. The resulting dimensionless problem is governed by two dimensionless parameters: the radii ratio $\eta = R_{in}/R_{out}$ and the Reynolds number $Re = \Omega R_{in} d/\nu$. Sometimes the Reynolds number is replaced by the Taylor number $Ta = 4\Omega^2 d^4/\nu^2$, whose definition also varies along numerous studies of this flow. The calculations were done by the finite volume method using 40×80×40 uniform grid in radial, azimuthal and axial directions, respectively.

It is well-known (see, e.g., [10,11]) that the first instability of the base one-dimensional Couette flow $(0, v_\theta(r), 0)$ results in steady axisymmetric Taylor vortices that, with further increase of the Reynolds number, undergo the second bifurcation and become oscillatory in time, and wavy in the azimuthal direction. For the following example, we consider the case with $\eta = 0.9$. According to [10], for this radii ratio, the flow becomes unstable at $Re_{cr} \approx 130$ (recalculated using definitions of [10]) with the axial period of Taylor vortices $\alpha = 3.2188$. For clarity, the flows below are visualized after the base state is withdrawn, and two axial periods are plotted.

In the course of calculations based on the standard finite volume discretization and fractional time step pressure / velocity decoupling, we observe steady axisymmetric Taylor vortices at $Re = 140$, which turn into time-periodic and wavy at $Re = 160$. At $Re = 200$ the time-dependence is non-periodic, and the vortices exhibit a quite complicated spatial pattern. Isosurfaces of the three velocity components are shown in Fig. 4, from which the axial symmetry at $Re = 140$, and waviness at $Re = 160$, are clearly seen. It is more complicated to understand the flow structure at $Re = 200$. Also, all the patterns of Fig.4, even at smaller $Re = 140$ and 160, do not represent the Taylor vortices as they are usually seen in experiments (see, e.g., [11]).

Isosurfaces of the vector potentials $\Psi^{(\theta)}, \Psi^{(r)}$ and $\Psi^{(z)}$ are shown in Fig. 5. The isosurfaces of $\Psi^{(\theta)}$ correspond to the divergence free projection of the velocity field on the coordinate planes $\theta = const$, i.e., on the meridional planes $(r, z)$. In the case of axisymmetric flow ($Re = 140$) cross-sections of $\Psi^{(\theta)}$ isosurfaces with the planes $(r, z)$ yield streamlines of the meridional part of the flow. In this case the isosurfaces attain shapes of axisymmetric tori. With increase of the Reynolds number we observe break of the axial symmetry ($Re = 160$) and appearance of azimuthal waviness. In particular, the isosurfaces "thickness" oscillates along the circumferential direction. At $Re = 200$ the time-dependence turns into non-periodic, while the difference between wide and narrow parts of the $\Psi^{(\theta)}$ isosurfaces becomes more profound. Comparing the first row of frames in Fig. 5 with experimental photos of the Taylor vortices (e.g., Fig, 1 and 6 of [11]), we observe clear similarity. This similarity is expected, since experiments visualize mainly meridional flow, which is described here by the vector potential $\Psi^{(\theta)}$.

Focusing now on two other potentials shown in Fig 5, $\Psi^{(r)}$ and $\Psi^{(z)}$, we observe that a break of the axial symmetry and further pattern change is seen only on isosurfaces of $\Psi^{(r)}$, while isosurfaces of $\Psi^{(z)}$ remain almost axisymmetric. This means that the symmetry breaking instability affects flow projections on the $\theta = const$ planes ($\Psi^{(\theta)}$) and $r = const$ cylindrical surfaces ($\Psi^{(r)}$), but motion along the axial cross-sections $z = const$ remains almost axisymmetric. A more detailed illustration of the latter is given in Figs. 6-8 showing zoomed in isosurfaces together with the corresponding vector fields.

In the case of axisymmetric flow ($Re = 140$), shown additionally in Fig. 6, we observe meridional motion tangent to the isosurfaces of $\Psi^{(\theta)}$, which, as is mentioned above, represents the meridional stream function. Projections on two other coordinate surfaces result in flows with non-zero $\theta$-velocity component and zero $r$- and $z$- components. Since in the axisymmetric flow case $v_r$ and $v_z$ are not dependent on $\theta$, the above projection procedure zeroes meridional components when projecting on the $r = const$ and $z = const$ planes, and results into one non-zero circumferential velocity component dependent on the meridional coordinates $r$ and $z$. The latter follows from the above iteration process. At the first iterations stage, the results of two-dimensional divergences $\nabla_{(r,\theta)} \cdot \boldsymbol{v}$ and $\nabla_{(\theta,z)} \cdot \boldsymbol{v}$ depend only on $r$ and $z$. Apparently, the solutions $\varphi$ of the corresponding two-dimensional Poisson equations are also $\theta$-independent. This means that the circumferential velocity component is not altered at the second iterations stage. Taking into account the no-slip boundary conditions, the resulting divergence free vector can have only zero radial and axial components.

Another conclusion of the above explanation is that the circumferential velocity component is the same when projecting on either $r = const$ or $z = const$ surfaces, which is also observed in the calculations. This component describes deviation of the circumferential velocity from its base state caused by convection of rotation in the meridional planes. The arrows shown on the isosurfaces of $\Psi^{(r)}$ and $\Psi^{(z)}$ in Fig. 6 correspond to the same azimuthal velocity, but are plotted on the cylindrical surfaces $r = const$ for $\Psi^{(r)}$, and on axial cross-sections $z = const$ for $\Psi^{(r)}$.

In the wavy flow at $Re = 160$ (Fig. 6), the isosurfaces of $\Psi^{(\theta)}$ and the corresponding arrow plots remain similar to those of the axisymmetric case (Fig. 6). The plotted pattern of $\Psi^{(r)}$ contains isosurfaces shown in the corresponding frame of Fig. 6, and an arrow plot done at the middle cylindrical surface. To get more insight in the $\Psi^{(z)}$ pattern, we plot a fragment of its several axial cross-sections, where we show isolines of $\Psi^{(z)}$ (colors) and an arrow plot of the projected velocity. The azimuthal dependence is clearly observed in the $\Psi^{(r)}$ pattern, but is not seen in the plot of $\Psi^{(z)}$. Also, the projected velocity vectors noticeably deviate from the circumferential direction on the plot of $\Psi^{(r)}$, but there is no visible deviation on the plot of $\Psi^{(z)}$. The latter may be only an optical effect. Comparison of maximal values of components of velocity projections shows that the maximal absolute values of circumferential components are 0.269 and 0.175 for projections on the $r = const$ and $z = const$ surfaces, respectively. The corresponding maximal absolute values of the axial and radial components of the two projections are 0.0279 and 0.00706. Thus, deviations from the circumferential direction are very small, but comparable.

At $Re = 200$, the azimuthal dependence of the meridional part of the flow, described by $\Psi^{(\theta)}$, becomes very strong. Three upper frames of Fig. 8 illustrate how the flow projected on the meridional planes moves around wide and narrow parts of the $\Psi^{(\theta)}$ isosurfaces. Patterns of $\Psi^{(r)}$ and $\Psi^{(z)}$ are plotted in the same way as in Fig. 7. Here we already clearly observe azimuthal dependence of isolines of $\Psi^{(z)}$ and the arrows deviate from the circumferential direction. Again, it is difficult to see any deviation from the circumferential direction on the $\Psi^{(r)}$ plot, but comparison of the projection components values show that the deviation remains comparable. Thus, the maximal absolute values of circumferential components are 0.350 and 0.316 for projections on the $r = const$ and $z = const$ surfaces, respectively. The corresponding maximal absolute values of the axial and radial components of the two projections are 0.0681 and 0.0225.

*5.3. Visualization of 3D-PTV experimental data in a shallow embayment*

For the third example, we visualize three-dimensional experimental data. The experiment [12] sketched in Fig. 9 was conducted in a shallow embayment that mimicked a similar flow in a river. The experimental "embayment" is 25 mm long (*x*-direction) and is 25 mm wide (*y*-direction), its depth is 3 mm (*z*-direction). The Reynolds number based on the water depth and the free-stream velocity is $Re = 3900$. In the course of the experiment, three components of the velocity field were measured on a three-dimensional grid with 31 nodes in the *x*- and *y*-directions and 11 nodes in the *z*-direction. The experimental data was then interpolated onto a $40^3$ staggered grid, after which the whole visualization procedure was carried out.

Examining the calculated vector potentials and the corresponding vector plots (Fig. 9b-9d), we see that the main flow circulation in the (*x,y*) planes (Fig. 9b) is represented rather well, and isosurfaces of $\Psi^{(z)}$ clearly describe its pattern. The two other vector potentials describe three-dimensional additions to the quasi-2D circulation of Fig. 9b. Comparing maximal and minimal values of the three potentials, we observe that the potential $\Psi^{(z)}$ maximal absolute value is more than an order of magnitude larger than that of the two other potentials, which are comparable. This shows that the main circulation carries most of the flow energy. In addition we observe that circulations in the (*y,z*) planes (Fig. 9c) are noticeable in the whole bulk of the flow and change in their direction is different for upstream and downstream values of *x*. The circulations in the (*x,z*) planes (Fig. 9d) are most intensive near embayment edges *y*=0 and 0.25, where the isosurfaces are plotted. There are also weaker circulations in the central part of the embayment, but to make them visible we had to plot arrows with uniform, not connected to the velocity magnitude, lengths. It should be emphasized, however, that the plots in Figs. 9c and 9d are based on 11 experimental points along the *z* direction, which can be insufficient for an accurate numerical post-processing. These two results are yet to be verified either by computation, or by more representative experimental data.

## 6. Conclusions

In this article we presented an alternative method for computation of quasi-two-dimensional velocity projections needed for visualization of three-dimensional incompressible flows, as it was proposed in our earlier paper [1]. The method is based on the SIMPLE-like iterations and removes all the previous restrictions on the shape of the flow region and system of coordinates. It is shown also that for the standard finite volume formulation on the staggered grid, the proposed approached may converge in a single iteration.

The proposed method is illustrated on three velocity fields calculated (measured) and visualized in Cartesian, cylindrical and spherical coordinates. In each case, the visualization allowed us to see some flow features that were not noticed on commonly used velocity or vorticity plots. This visualization approach is suitable for visualization of not only numerical, but also experimental data, as is shown by the last example.

**Acknowledgement**

The author wishes to express his gratitude to Yuri Feldman and Yulia Akutina for providing their numerical and experimental data that was used in the above examples.

**Figure captions**

Figure 1. Visualization of convective flow of air from a hot sphere placed in the center of a cold cubic box at $Ra=10^4$ (upper frames and $Ra=10^6$ (lower frames). Translucent red surface shows border of the sphere. Isosurfaces of $\Psi^{(y)}$, $\Psi^{(x)}$ and $\Psi^{(z)}$ are shown by colors and are superimposed with the corresponding vector plots of quasi-two-dimensional projections of velocity fields.

Figure 2. Visualization of convective flow of air from a hot sphere placed in the center of a cold cubic box at $Ra=10^4$ (upper frames and $Ra=10^6$ (lower frames). Isosurfaces of $\Psi^{(\phi)}$, $\Psi^{(\theta)}$ and $\Psi^{(R)}$ calculated after the data was transformed from the Cartesian grid to the spherical coordinates.

Figure 3. Visualization of convective flow of air from a hot sphere placed in the center of a cold cubic box. A characteristic isosurface of (a) $\Psi^{(\phi)}$, (b) $\Psi^{(R)}$, and (c) $\Psi^{(\theta)}$ with the corresponding vector plots of quasi-two-dimensional projections of velocity fields on the coordinate surfaces $\phi = const, R = const, \theta = const$, respectively.

Figure 4. Isosurfaces of velocity components of the Taylor – Couette flow at different Reynolds numbers. The radii ratio is 0.9, the axial period is 2.0082. Two axial periods are shown.

Figure 5. Isosurfaces of vector potentials of the Taylor – Couette flows shown in Fig. 4.

Figure 6. Fragments of vector potentials of the Taylor – Couette flow at $Re=140$ with vector plots of the corresponding velocity projections.

Figure 7. Fragments of vector potentials of the Taylor – Couette flow at $Re=160$ with vector plots of the corresponding velocity projections.

Figure 8. Fragments of vector potentials of the Taylor – Couette flow at $Re=200$ with vector plots of the corresponding velocity projections.

Figure 9. Post processing of experimental results on the flow in a shallow embayment. (a) Sketch of the experimental setup. (c)-(d) Vector potentials and the corresponding arrow plots calculated using experimental 3D-PTV data.

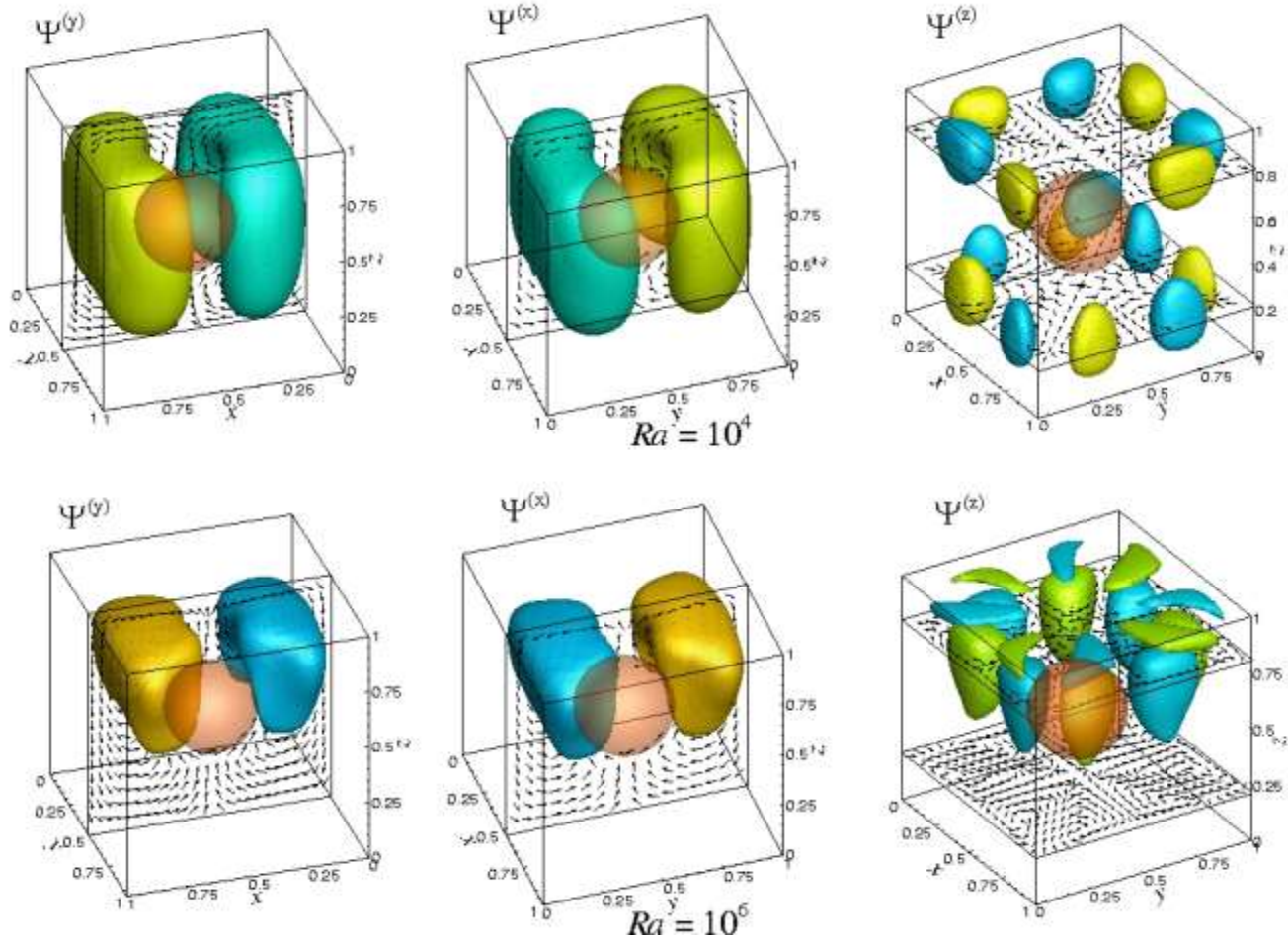

|  | $Ra=10^4$ | $Ra=10^6$ |
|---|---|---|
| $\min \Psi^{(y)}, \max \Psi^{(y)}$ | ±0.0177 | ±0.029 |
| $\min \Psi^{(x)}, \max \Psi^{(x)}$ | ±0.0177 | ±0.029 |
| $\min \Psi^{(z)}, \max \Psi^{(z)}$ | ±0.00099 | ±0.0026 |
| levels of $\Psi^{(y)}$ | ±0.006 | ±0.015 |
| levels of $\Psi^{(x)}$ | ±0.006 | ±0.015 |
| levels of $\Psi^{(z)}$ | ±0.0005 | ±0.001 |

Figure 1. Visualization of convective flow of air from a hot sphere placed in the center of a cold cubic box at $Ra=10^4$ (upper frames and $Ra=10^6$ (lower frames). Translucent red surface shows border of the sphere. Isosurfaces of $\Psi^{(y)}$, $\Psi^{(x)}$ and $\Psi^{(z)}$ are shown by colors and are superimposed with the corresponding vector plots of quasi-two-dimensional projections of velocity fields.

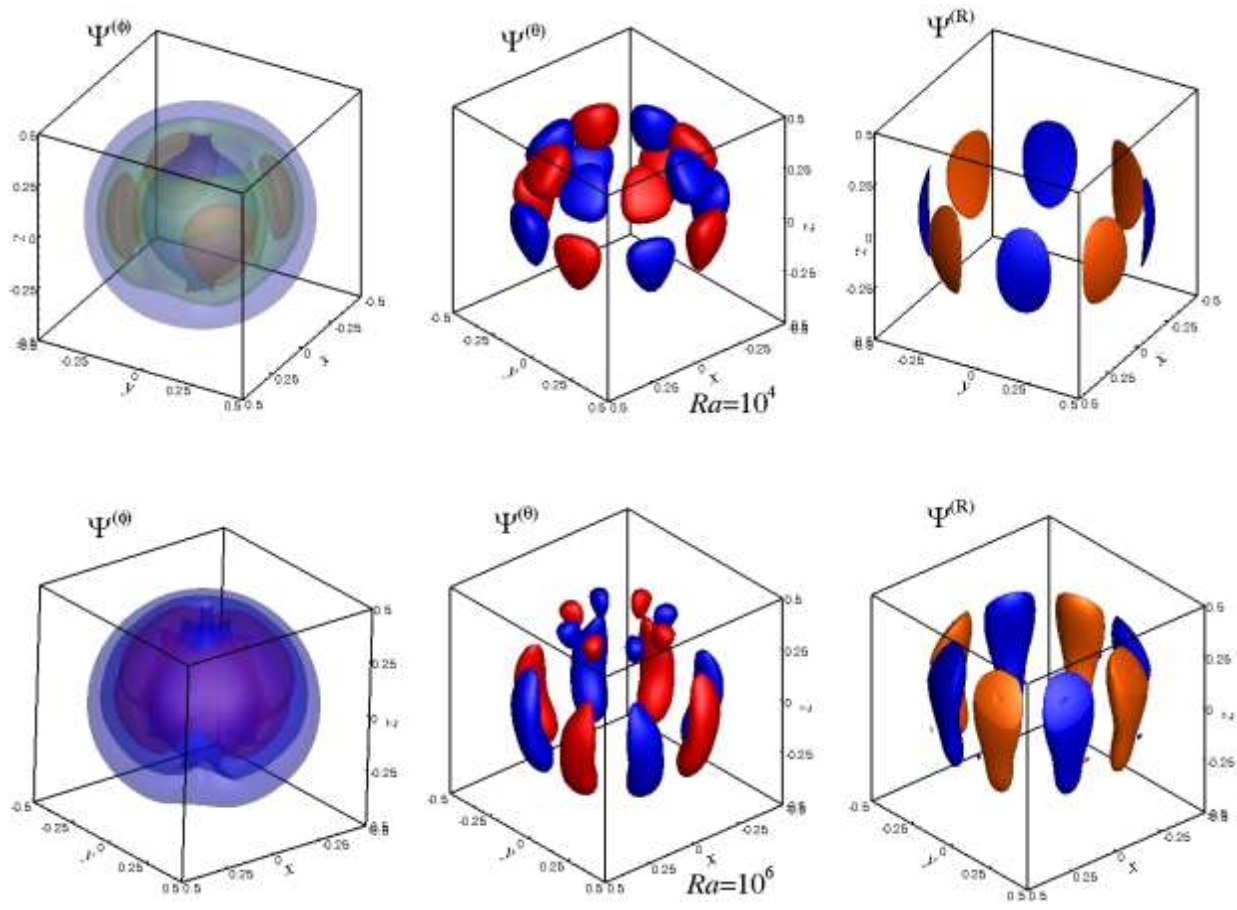

|  | $Ra=10^4$ | $Ra=10^6$ |
|---|---|---|
| $\min \Psi^{(\phi)}, \max \Psi^{(\phi)}$ | 0, 0.0179 | 0, 0.0224 |
| $\min \Psi^{(\theta)}, \max \Psi^{(\theta)}$ | ±0.000227 | ±0.000606 |
| $\min \Psi^{(R)}, \max \Psi^{(R)}$ | ±0.00523 | ±0.00439 |
| levels of $\Psi^{(\phi)}$ | 0.002, 0.01, 0.015 | 0.003, 0.01, 0.015 |
| levels of $\Psi^{(\theta)}$ | ±7·10⁻⁵ | ±6·10⁻⁴ |
| levels of $\Psi^{(R)}$ | ±0.003 | ±0.002 |

Figure 2. Visualization of convective flow of air from a hot sphere placed in the center of a cold cubic box at $Ra=10^4$ (upper frames and $Ra=10^6$ (lower frames). Isosurfaces of $\Psi^{(\phi)}$, $\Psi^{(\theta)}$ and $\Psi^{(R)}$ calculated after the data was transformed from the Cartesian grid to the spherical coordinates.

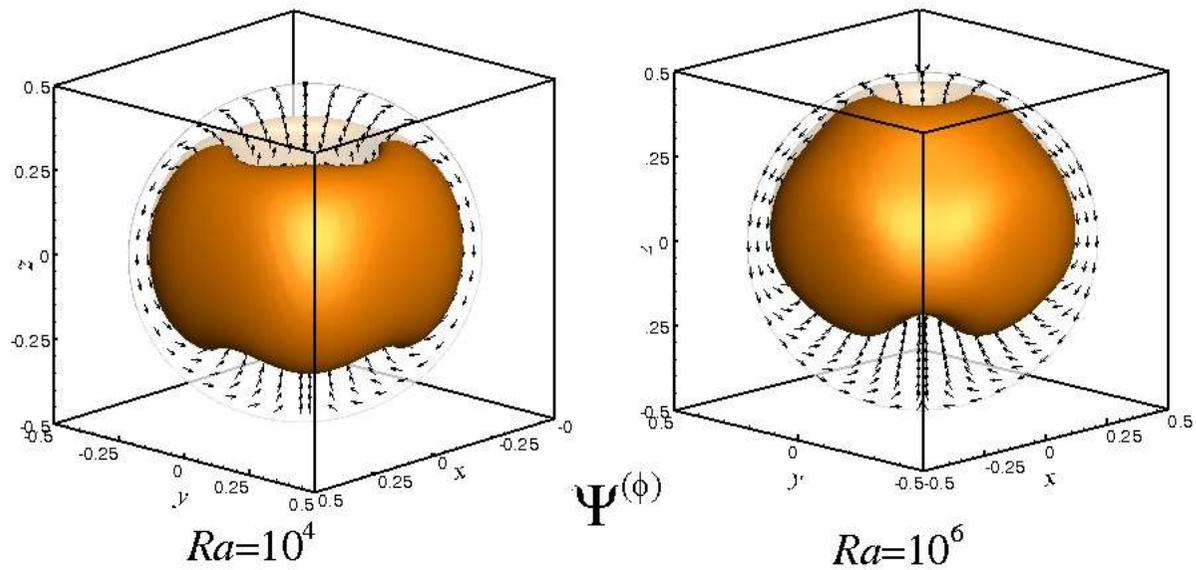

Figure 3a. Visualization of convective flow of air from a hot sphere placed in the center of a cold cubic box. A characteristic isosurface of $\Psi^{(\phi)}$ with the corresponding vector plots of quasi-two-dimensional projections of velocity fields on the coordinate surfaces $\phi = 0$ and $\pi$.

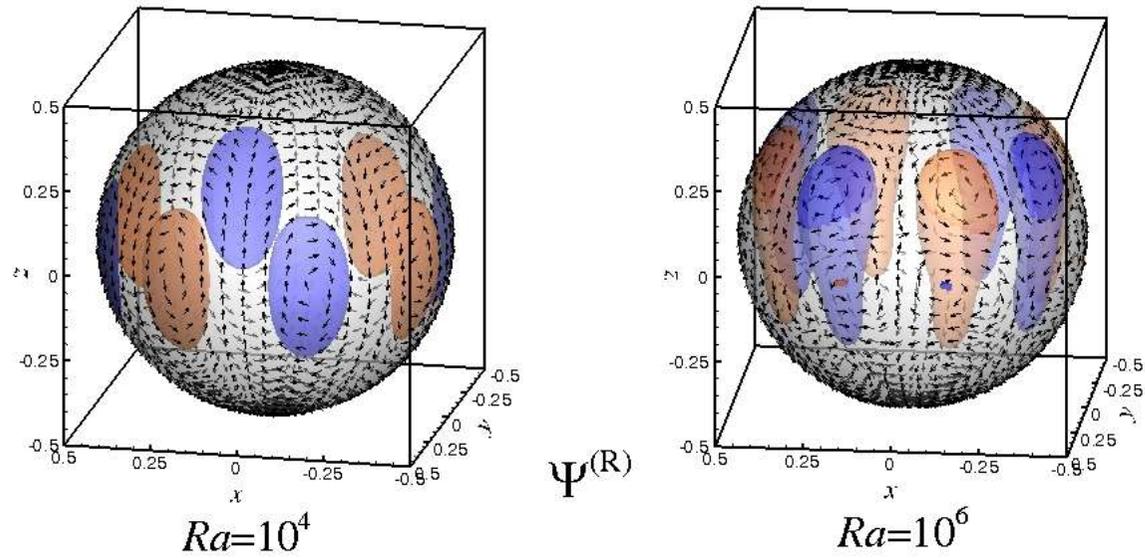

Figure 3b. Visualization of convective flow of air from a hot sphere placed in the center of a cold cubic box. A characteristic isosurface of $\Psi^{(R)}$ with the corresponding vector plots of quasi-two-dimensional projections of velocity fields on the coordinate surfaces $R = const$.

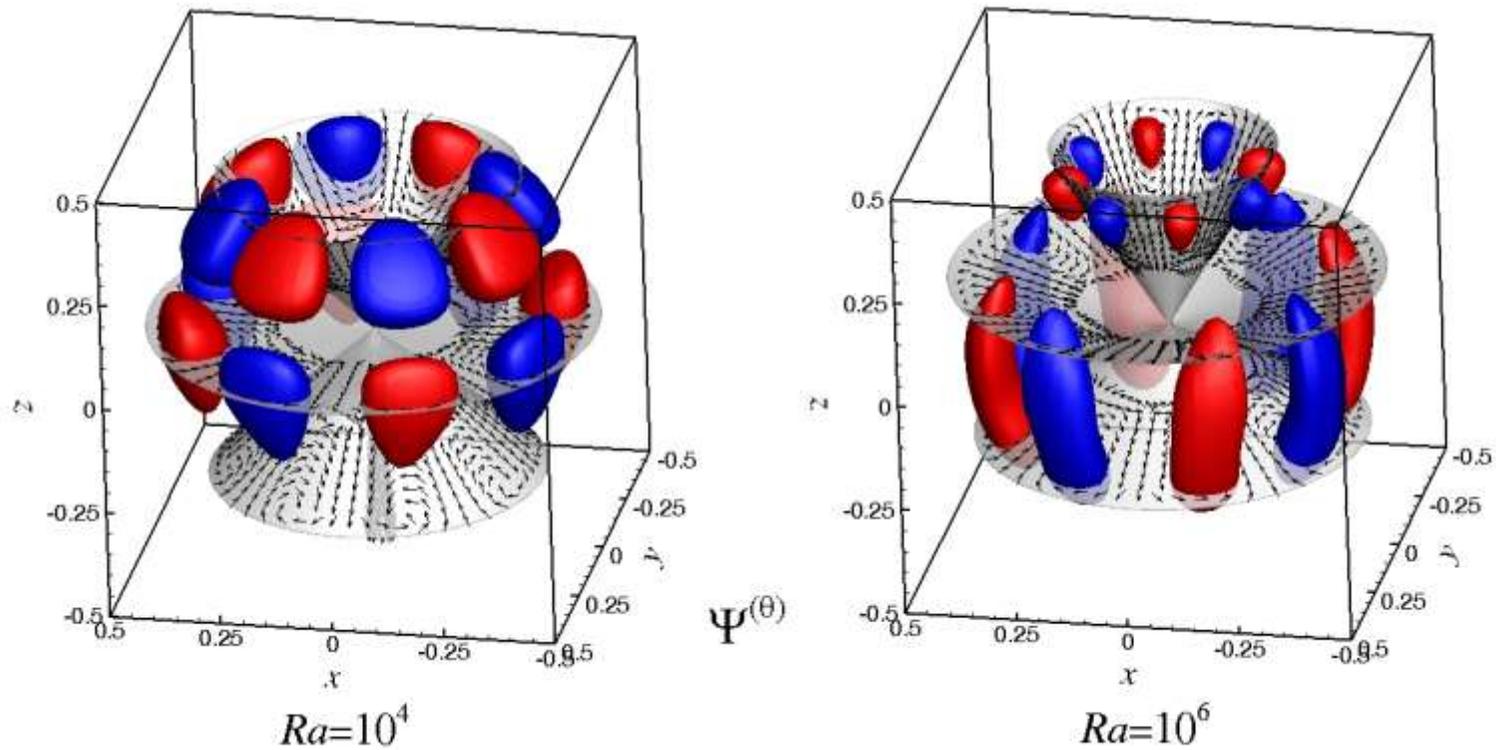

Figure 3c. Visualization of convective flow of air from a hot sphere placed in the center of a cold cubic box. A characteristic isosurface of $\Psi^{(\theta)}$ with the corresponding vector plots of quasi-two-dimensional projections of velocity fields on the coordinate surfaces $\theta = const$.

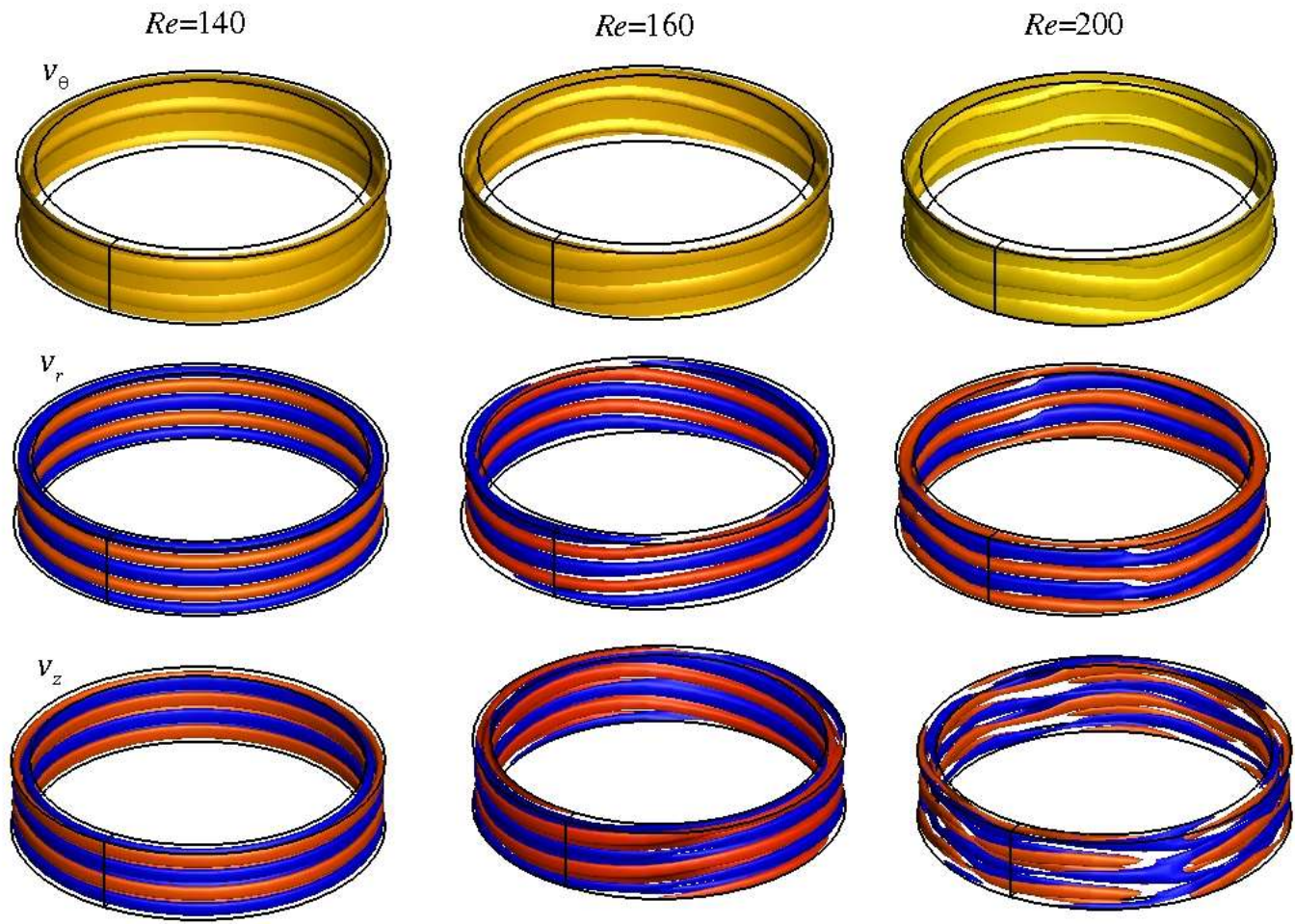

|  | $Re$=140 | $Re$=160 | $Re$=200 |
|---|---|---|---|
| min $v_\theta$, max $v_\theta$ | 0, 1 | 0, 1 | 0, 1 |
| min $v_r$, max $v_r$ | -0.0253, 0.0353 | -0.0398, 0.0607 | -0.0612, 0.0906 |
| min $v_z$, max $v_z$ | ±0.0322 | ±0.0561 | ±0.0946 |
| levels of $v_\theta$ | 0.25 | 0.25 | 0.25 |
| levels of $v_r$ | ±0.01 | ±0.015 | ±0.02 |
| levels of $v_z$ | ±0.01 | ±0.02 | ±0.04 |

Figure 4. Isosurfaces of velocity components of the Taylor – Couette flow at different Reynolds numbers. The radii ratio is 0.9, the axial period is 2.0082. Two axial periods are shown.

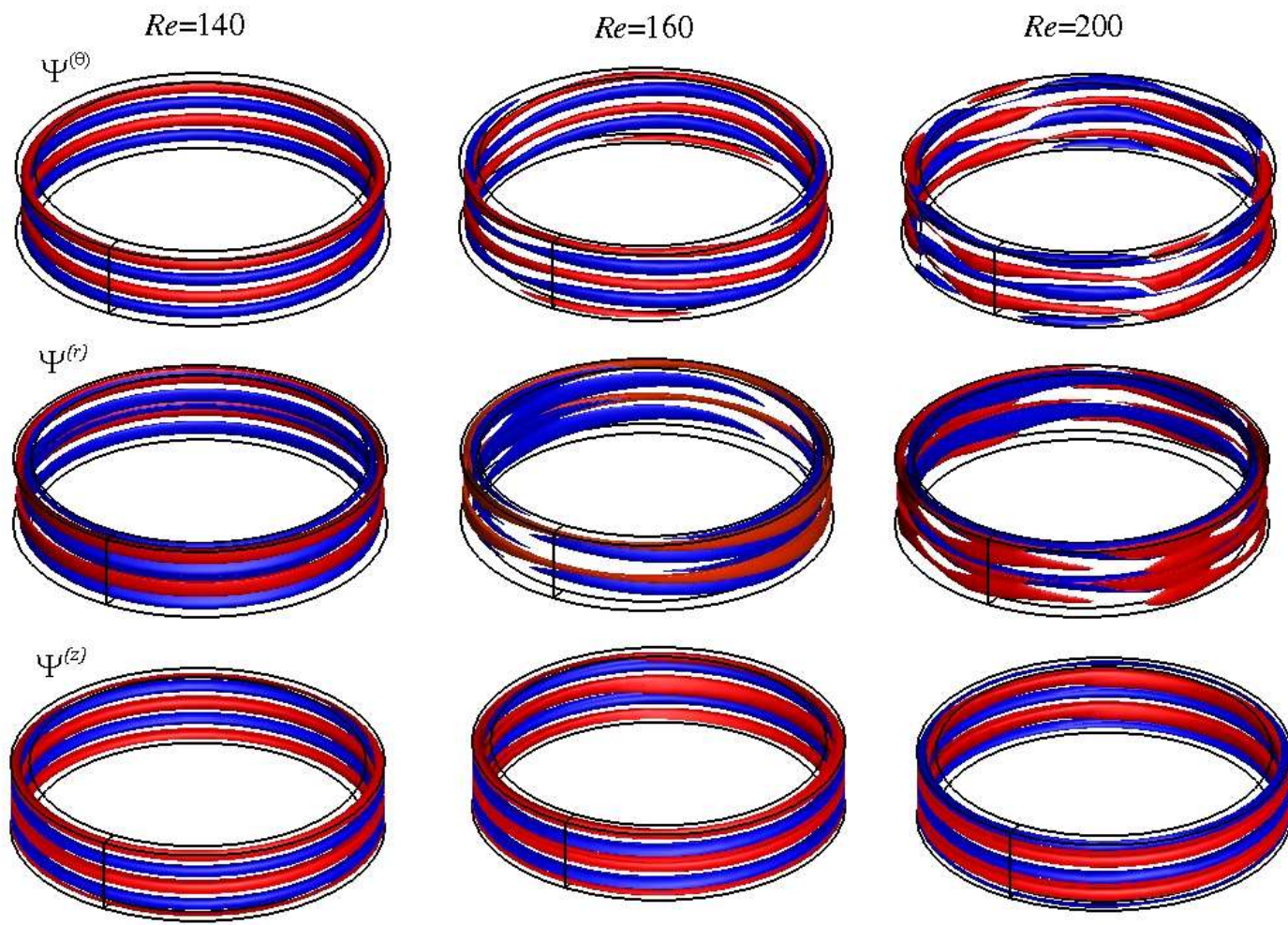

| | $Re$=140 | $Re$=160 | $Re$=200 |
|---|---|---|---|
| $\min \Psi^{(\theta)}, \max \Psi^{(\theta)}$ | ±0.00975 | ±0.0185 | ±0.0309 |
| $\min \Psi^{(r)}, \max \Psi^{(r)}$ | -0.0615, 0.0815 | -0.304, 0.139 | -0.316, 0.358 |
| $\min \Psi^{(z)}, \max \Psi^{(z)}$ | -0.107, 0.0983 | -0.304, 0.139 | -0.316, 0.358 |
| levels of $\Psi^{(\theta)}$ | 0.005 | 0.008 | 0.012 |
| levels of $\Psi^{(r)}$ | -0.03, 0.04 | -0.15, 0.06 | ±0.12 |
| levels of $\Psi^{(z)}$ | ±0.05 | -0.02, 0.03 | -0.05, 0.06 |

Figure 5. Isosurfaces of vector potentials of the Taylor – Couette flows shown in Fig. 4.

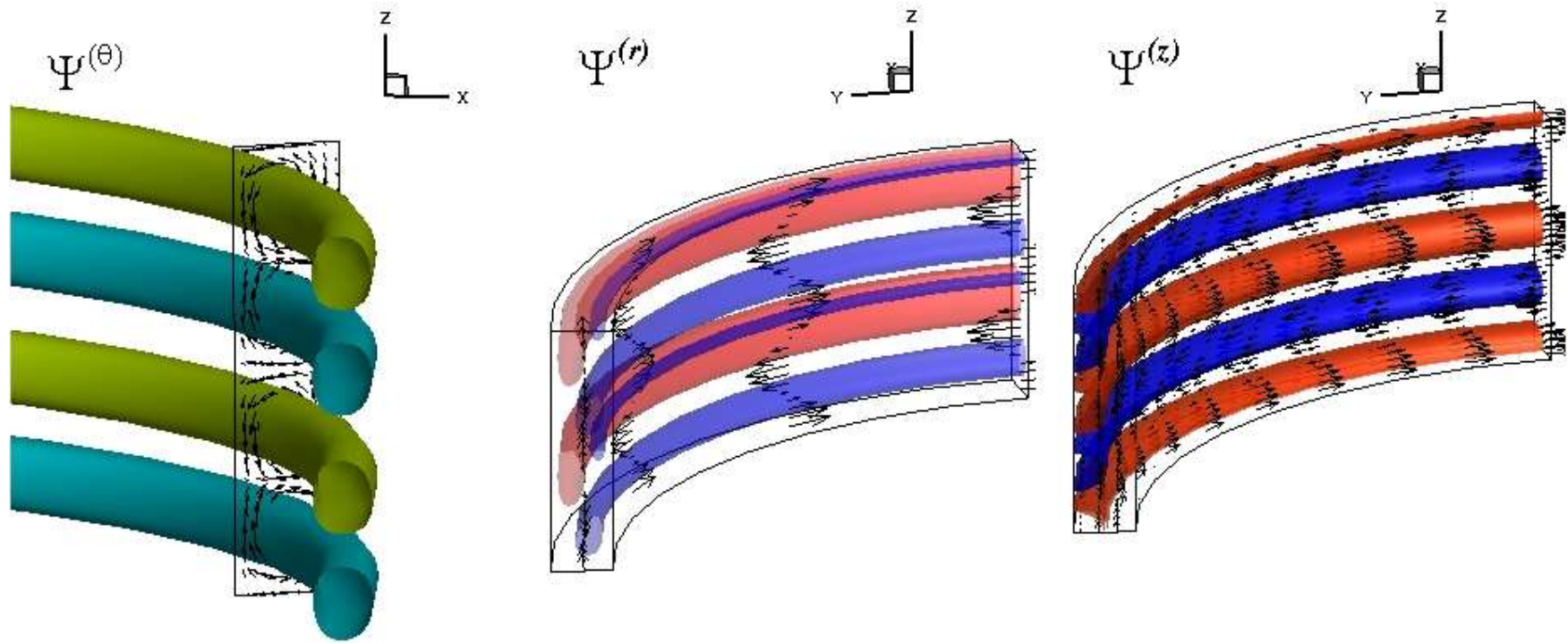

Figure 6. Fragments of vector potentials of the Taylor – Couette flow at *Re*=140 with vector plots of the corresponding velocity projections.

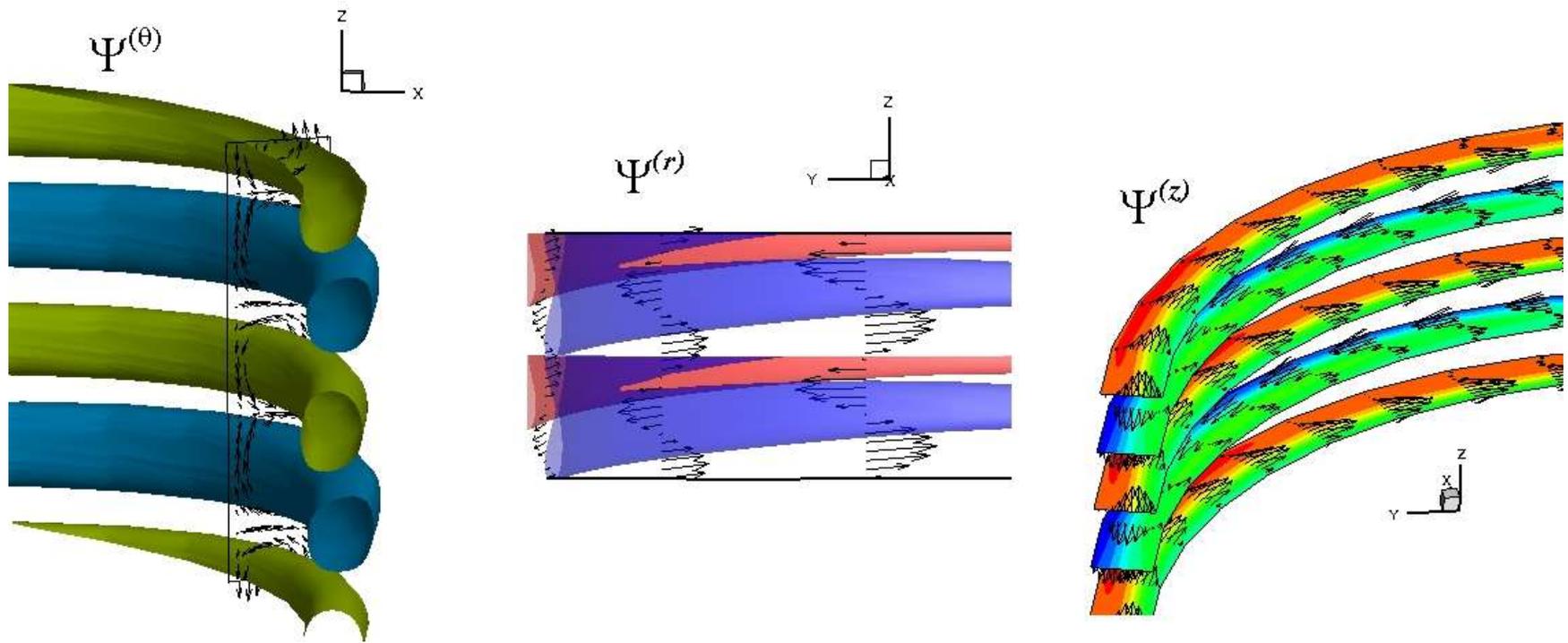

Figure 7. Fragments of vector potentials of the Taylor – Couette flow at *Re*=160 with vector plots of the corresponding velocity projections.

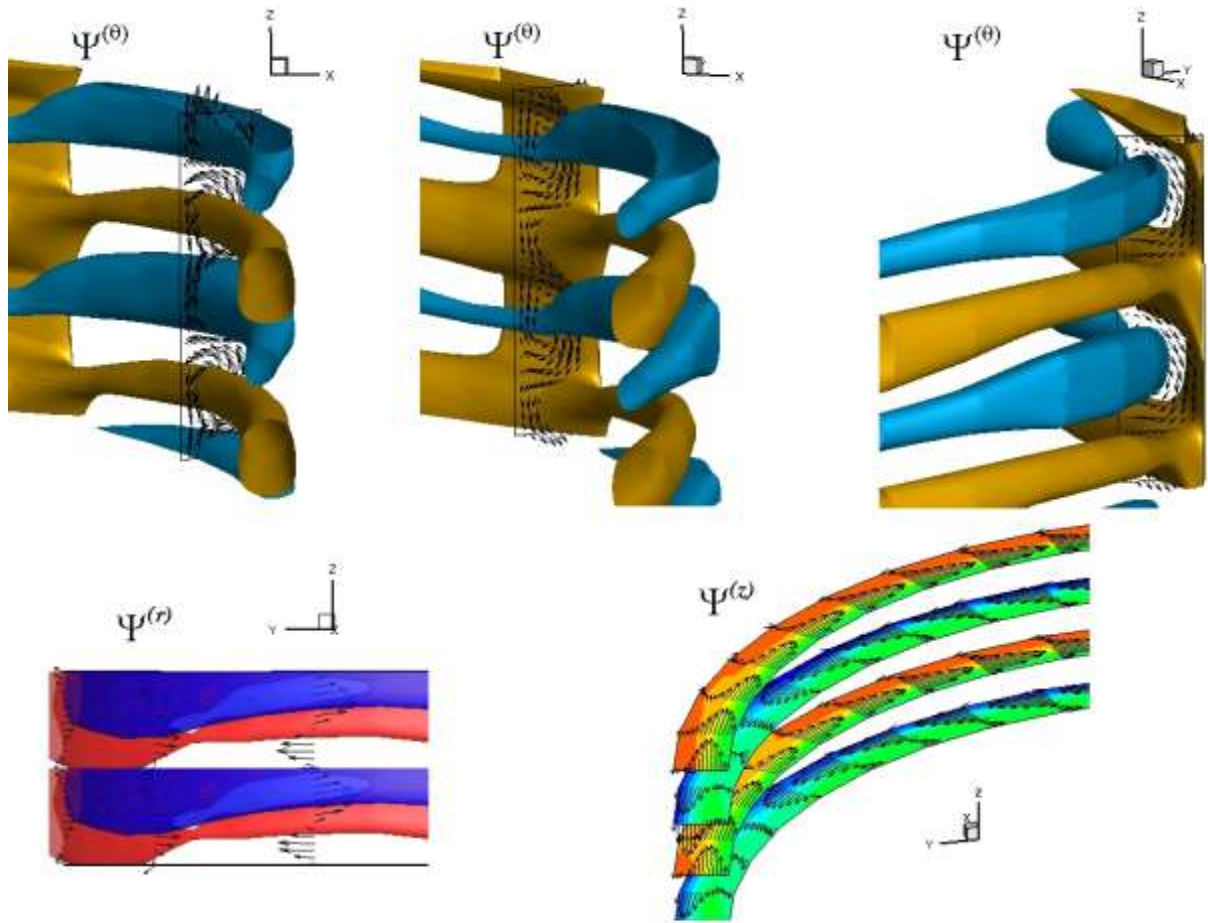

Figure 8. Fragments of vector potentials of the Taylor – Couette flow at *Re*=200 with vector plots of the corresponding velocity projections.

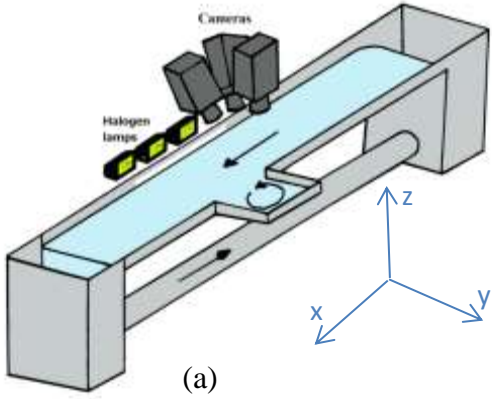
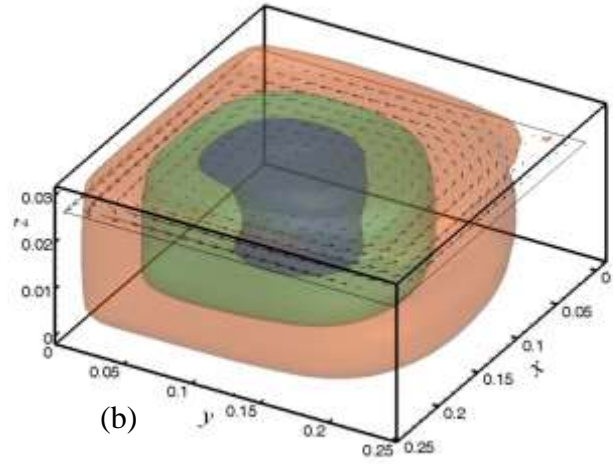

(a)

(b)

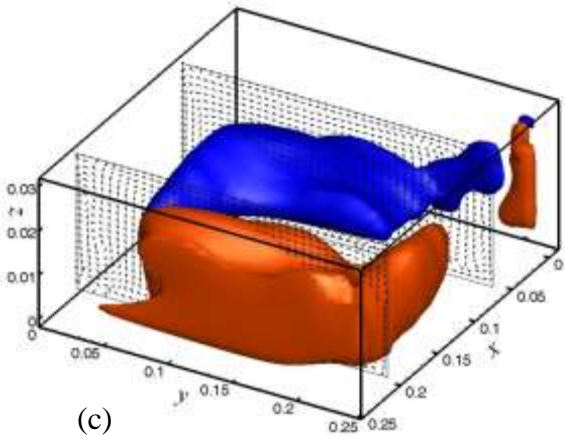
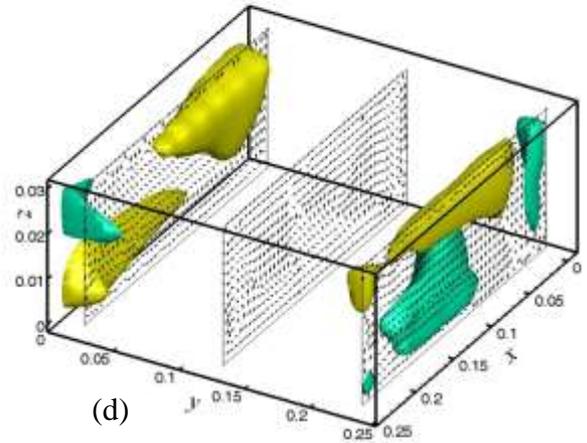

(c)

(d)

| min $\Psi^{(x)}$, max $\Psi^{(x)}$ | -0.00026, 0.00060 |
| --- | --- |
| min $\Psi^{(y)}$, max $\Psi^{(y)}$ | -0.00035, 0.00041 |
| min $\Psi^{(z)}$, max $\Psi^{(z)}$ | -0.0081, 0.00049 |
| levels of $\Psi^{(x)}$ | ±0.0001 |
| levels of $\Psi^{(y)}$ | -0.0001, 0.00015 |
| levels of $\Psi^{(z)}$ | -0.0008, -0.005, -0.007 |

Figure 9. Post processing of experimental results on the flow in a shallow embayment. (a) Sketch of the experimental setup. (c)-(d) Vector potentials and the corresponding arrow plots calculated using experimental 3D-PTV data.